\documentstyle[psfig]{article}
\begin{document}

\noindent {\Large {\bf Squeezing with cold atoms}}

\vspace{0.5cm}

{\sc A. Lambrecht, T. Coudreau, A.M. Steinberg and E. Giacobino,} \\

{\it Laboratoire Kastler Brossel, UPMC, ENS et CNRS,}

{\it Universit\'e Pierre et Marie Curie, case 74, F75252 Paris, France} \\

({\sc EuroPhysics Letters} {\bf 36}, p. 93 (1996))\\

PACS: 32.80.Pj; 42.50.Lc; 42.65.Pc \\

\noindent {\bf Abstract. -}
Cold atoms from a magneto-optic trap have been used as a nonlinear 
($\chi^{(3)}$) medium in
a nearly resonant cavity. Squeezing in a probe beam passing through the
cavity was demonstrated. The measured noise reduction is 40\% for free atoms
and 20\% for weakly trapped atoms.

\vspace{0.5cm}

Soon after the first implementations of magneto-optic traps \cite{ref1a,%
ref1b}, the strong nonlinear properties of laser cooled atoms were
recognized. It was shown that a probe beam going through a cloud of cold
atoms could experience gain due to Raman transitions involving the trapping
beams \cite{ref2a,ref2b}. When the atoms were placed in a resonant optical
cavity, laser action corresponding to that gain feature was demonstrated 
\cite{ref3}. Meanwhile, when cold atoms are driven by a slightly
detuned probe laser beam inside an optical cavity, bistability is
observed at very low light powers \cite{ref4a}. This high nonlinear
dispersion comes from the fact that, since the atoms are virtually
motionless, the Doppler width is smaller than the natural linewidth and the
probe frequency can be set close to atomic resonance.
Such  nonlinear behavior indicates that the system is capable of
significantly modifying the quantum fluctuations of a probe beam, 
in particular leading to squeezing.
The generation of squeezed light through interaction with nonlinear
media has been subject of extensive theoretical and experimental 
studies (see for example \cite{ApplB}). The use of atomic media looked
particularly promising in the absence of Doppler broadening
\cite{castelli,reid,hilico};
laser-cooled atoms should
therefore be ideal. However, no experiment
performed so far has used cold atoms for squeezing, while those which
relied on atomic beams failed to live up to expectations.

In this paper we
report quantum noise reduction in a probe beam that has interacted 
with cold atoms 
in an optical cavity. Quadrature squeezing as large as $40\%$
(uncorrected for detection efficiency) was
measured at the output of the cavity. This value is the largest ever
measured in an atomic medium. 
Furthermore it is in good agreement with the
theoretical predictions. It therefore has very good prospects for 
leading to much higher levels of squeezing when the apparatus is 
suitably improved.
The squeezing was first observed in a free
cloud, just after the trap had been turned off. It was also observed
in a weakly bound trap with trapping beams turned down by a factor of $10$
as compared to the original trap.
In light of the rapid development of atom-cooling
technology, it is conceivable that a set-up of this type be constructed
on a much smaller scale, relying exclusively on diode lasers.  This
could lead to the realization of a compact and efficient quantum
``noise eater''.

Up to now, all experiments of this type had been performed
on atomic beams, where the transverse Doppler effect is small but
often not completely negligible\cite{raizen,hope,poizat}.
The best value of quantum noise reduction measured in that 
case is of the order 
of 20$\%$, while theory predicts higher figures that have never been obtained.

In contrast to atomic beams, cold atoms constitute a well controlled medium,
where theoretical models can in principle be fully tested against
experimental results, including realistic conditions such as the spatial 
character of the laser beam and additional noise sources.
In such conditions one is able to accurately model the
squeezing data theoretically for the first time, whereas in the atomic 
beam experiments no quantitative explanation has been given of the discrepancy
between the measured squeezing and the significantly higher theoretical
predictions.

To fit our experimental data, we have developped a theoretical model
which fully takes into account the transverse structure of the probe beam 
\cite{ref10}
and which also considers the influence of additional noise sources like atomic
number fluctuations \cite{Lambrecht}. With this treatment we have not only 
obtained a correct prediction for the magnitude of the quadrature squeezing,
but also a continuous prediction for the minimal and maximal noise power
as a function of cavity detuning, which fits the experimental result well.

The experimental set-up used to demonstrate squeezing with cold
atoms has been described in detail elsewhere 
\cite{ref5}. Here, we will only recall its main features. 
A circularly polarized probe beam is sent into a resonant
optical cavity containing a cloud of cold Cesium atoms prepared in a standard
magneto-optic trap\cite{ref1a,ref1b}. With large and rather intense trapping beams, 
detuned by $3$ times the linewidth below resonance
of the $6S_{1/2},\ F=4$ to $6P_{3/2},\ F=5$
transition, we obtain a cloud of $1$cm
in diameter, with densities on the order of $10^9$ atoms/cm$^3$. The
temperature of the atoms is of the order of some mK. The number of
atoms interacting with the probe beam is measured with a method described below.

The cavity is a $25$cm-long linear cavity, with a waist of $260\mu$m, built
around the cell (fig.1). The input mirror has a transmission coefficient of $10\%$
and the end mirror is highly reflecting. The cavity is thus close to
the ``bad cavity'' case, where the linewidth of the cavity ($5$MHz) is
larger than the radiative linewidth ($\gamma =2.6$MHz),
and which is expected to be the most favorable for squeezing. The cavity
is in the horizontal symmetry plane of the trap, making a $45^{\circ }$ angle with the
two trapping beams propagating in this plane. The probe beam, generated
by the same Ti:Sapphire laser as the trapping beams, can be detuned by $0$ to $130$MHz on 
either side of the $%
6S_{1/2},\ F=4$ to $6P_{3/2}\ ,F=5$ transition. We measure the
probe beam intensity transmitted through the cavity with a photodiode
located behind the end mirror.
\begin{figure}
\centerline{\psfig{figure=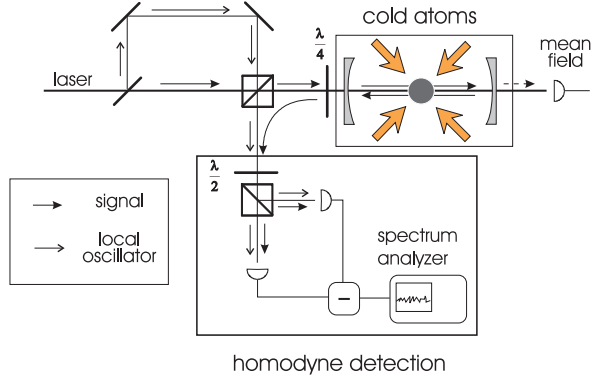,height=5cm}}
\caption{Experimental set-up designed to study quantum
fluctuations of a probe beam that has interacted with cold atoms in
a nearly single-ended cavity. PBS, polarizing beamsplitter; QWP, quarterwave plate;
HWP, halfwave plate; PD, photodiode.}
\end{figure}

The field coming out of the cavity is separated from the incoming one by an 
optical circulator,
made of a polarizing beamsplitter and a quarter-wave plate, and mixed with a 
local oscillator beam using the second
input port of the same beamsplitter.
Orthogonally polarized at the output of
the beamsplitter, the signal and local oscillator beams are split into
equal-intensity sum and difference fields by a half-wave plate and a
second polarizing beam splitter.
Both parts    
are detected by photodiodes with a quantum efficiency of $96\%$.
The total homodyne efficiency is of the order of $90\%$.
The ac parts of the photodiode currents are amplified and subtracted. The
resulting signal is further amplified and sent to a spectrum analyzer and to
a computer.

When scanning the cavity resonance, bistability due to the nonlinear dispersion of the cold 
atoms is easily
observed with incident powers as low as a few $\mu$W, as soon 
as the number $N$ of interacting
atoms is large enough. The nonlinear phase shift of the cavity giving rise to bistability 
is proportional to the cooperativity 
parameter $C$ also called bistability parameter: ($C=g^2N/\gamma T$, where $g$ is the 
atom-light coupling coefficient, $\gamma $
the radiative linewidth of the transition and $T$ the energy
transmission coefficient of the mirror). $C$ is determined for each recording by 
measuring the ratio between the amplitudes of the bistability curve and of the resonance 
curve of the empty cavity.
We find a cooperativity of the order of $%
100$ in presence of the trap. It should be noted that the atoms
are partly saturated by the trapping beams, and that therefore the measured $C$
value is smaller than the one corresponding to the
total number of atoms in the interaction area.

The fluorescence emitted by atoms excited by the trapping beams
constitutes a source of excess noise, which decreases the
amount of quantum noise reduction attainable in principle.  
To avoid this
effect, we turn off the trap during the noise measurements. With this
method the measurement time is limited
to $20$ to $30$ms, due to expansion and free fall of the atomic
cloud.
When the trap is turned off, the number of interacting atoms
becomes time-dependent. The variation of the refractive
index due to the escape of the atoms out of the probe beam provides a
natural scan of the cavity across resonance. In this way, the resonance
peak is scanned in about $10$ms. However, under these conditions, 
the $C$ value, being proportional to $N$, is no longer constant over the scan and it
becomes necessary to adopt a specific model of its time-dependence
in order to interpret the noise spectra.
A model for the variation of $C$ with time can be obtained by calculating
the variation of the linear phase shift caused by an expanding and falling
ensemble of atoms in a Gaussian laser beam \cite{Lambrecht}.

The atomic sample is assumed to have initially a Gaussian velocity and
position distribution.
Supposing according to experimental conditions
the Rayleigh length of the beam to be much larger than the cloud size 
which itself is
large compared to the beam waist, the variation of $%
C\left( t\right) $ is given by the product of a Lorentzian function 
representing the ballistic flight of the atoms with an exponential function
accounting for the effect of gravity:
\begin{equation}
C\left( t\right) =C\left( 0\right) \frac{\tau _r^2}{\tau _r^2+t^2}\exp
\left( -\frac{t^4}{\tau _g^2\left( \tau _r^2+t^2\right) }\right)
\label{Coop}
\end{equation}
$C\left( 0\right) $ is the cooperativity value right after the
trap is turned off; the time constant $\tau _r=\sigma _r/\sigma
_v $ is the time for the atoms with temperature $T_{\rm at}$ and with mean velocity $%
\sigma _v=\sqrt{kT_{\rm at}/m}$ to fly through the cloud of radius $\sigma _r$ and $%
\tau _g=2\sqrt{2}\sigma _v/g$ is the time it takes for the falling atoms to
accelerate to $2\sqrt{2}$ times their original thermal velocity.

To check this model, the cooperativity was experimentally studied as a
function of time. 
fig.2 shows the result of such a measurement.
The experimental points are fitted by the theoretical expression (\ref{Coop}). 
The line correponds to the theoretical prediction for 
a cloud radius of $4$mm and a
temperature of $5$mK, which are 
consistent with the characteristics of our trap.
As the theoretical curve fits the experimental data very well, these
parameters were subsequently used to calculate the noise spectrum.
\begin{figure}
\centerline{\psfig{figure=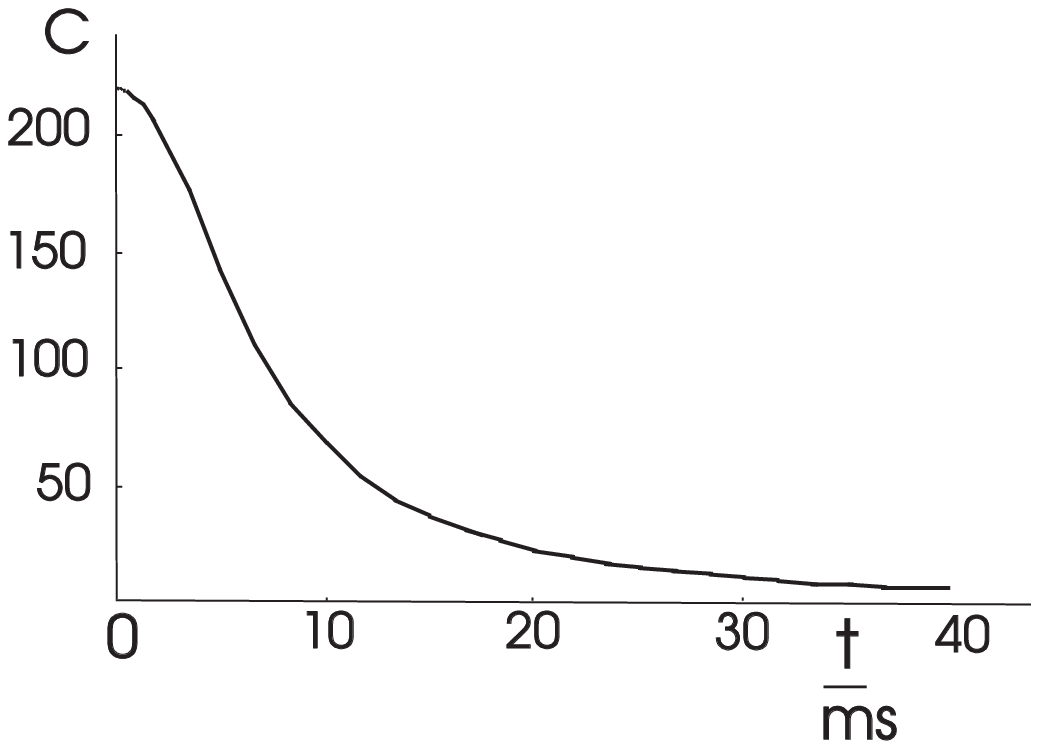,height=5cm}}
\caption{Time dependence of the cooperativity parameter $C$ as the
cloud of cold atoms is released at $t=0$. The dots correspond to experimental
data as described in the text, the curves are derived from Eq.(1).}
\end{figure}

While the cavity resonance is
scanned by the escape of the atoms the field fluctuations of the output 
beam are monitored with the homodyne
detection described above at a fixed analyzing frequency of $5$MHz. 
At the same time,
the local oscillator phase is rapidly varied with a piezoelectric
transducer to explore all noise quadratures of the probe beam.
As we have only $20-30$ms for the measurement, the phase of the 
local oscillator
must be scanned at frequencies on the order of kHz, which determines the analyzing
bandwidth of the spectrum analyzer to be about $100$kHz. 
The video bandwidth of
the spectrum analyzer should be adjusted to avoid any distortion
of the spectrum. As the videofilter of our model (Tektronix 2753 P)
did not have enough flexibility, we used numerical filters in the
processing of the spectrum.

The spectrum shown in fig.3 was obtained in such a manner.
The electronic noise was substracted from the signal provided by the spectrum analyzer 
before the filtering process. The averaged
shot noise level (determined by blocking the cavity) is indicated by the 
straight line. It can be seen that the noise
on the left hand side of the figure, when the cavity is out of resonance, is
at shot noise, whereas it goes below shot noise on the lower branch of the
bistability curve. The observed squeezing is $(40 \pm 10)\%$. On the upper
branch (right hand part of the figure), large excess noise is observed in
some quadratures. The powers of the probe laser and local oscillator were $%
25\mu$W and $9$mW.
The large ratio of these powers ensures the noise measured by the homodyne detection does 
not require any normalization, even if the probe beam reflected by the cavity does not have 
a constant intensity. The detuning from atomic resonance was $52$MHz below
resonance. The cooperativity parameter $C$ was found to be $220$ right after
turning off the trap. The error bar for the squeezing measurement
is due to the width of the random noise of the signal which is generated
within the measurement system itself. This random noise varies
as the mean noise level and is estimated from the shot noise
level to be $\pm 10\%$ at the minimum of fig.3. The variation of the 
transmitted field reproduced in the insert of fig.3 shows that the system is 
slightly
below the bistability threshold. The spectrum presented in fig. 3 shows the best squeezing 
value obtained in a series of experiments where several parameters such as the detuning of 
the probe field from the cavity and atomic resonances, the analyzing frequency and the 
atomic number were varied. 
\begin{figure}
\centerline{\psfig{figure=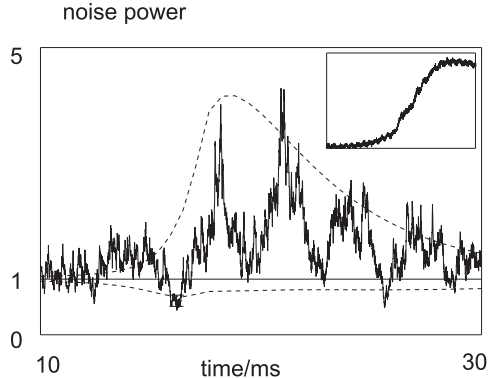,height=5cm}}
\caption{Noise signal taken with free atoms at a fixed
observation frequency of 5MHz
as a function of time, while the cavity resonance is
scanned by the departure of the atoms.
Oscillations correspond to phase scan of the local oscillator. 
The broken lines are theoretical predictions
for the minimal and maximal noise, the solid line indicates the shot noise level.
The probe beam was detuned by 52MHz below resonance and its
power was 25$\mu$W.
The error bar is due to the width of the
noise trace. The insert shows the corresponding mean intensity.}
\end{figure}

In a second series of experiments, we have looked for quantum noise reduction
in presence of the trap. As mentioned above, intense trapping
beams produce too much excess noise to allow observation of
squeezing. However, we have been able to find a compromise by first trapping
the atoms from the vapor with intense laser beams, and then
turning down the trapping beams to about $1/10$ of their original power. 
Under these conditions, the cooperativity parameter is on the order of $20$. The
noise spectrum shown in fig.4 is recorded under similar conditions as above
(detuning of 52MHz and incident probe power of $16\mu$W) except that the cavity is scanned 
by means of a
piezoelectric transducer while the number of atoms is constant during the measurement. The 
best value for squeezing obtained under such
conditions is on the order of $(20 \pm 10)\%$.
Here, the quantum noise reduction appears on the upper branch of the
bistability curve.

An interesting feature of this result is that the
steep edge on the left hand side of the bistability curve 
(upper trace in fig.4) is due to optical pumping,
rather than to saturation which is responsible for squeezing.
\begin{figure}
\centerline{\psfig{figure=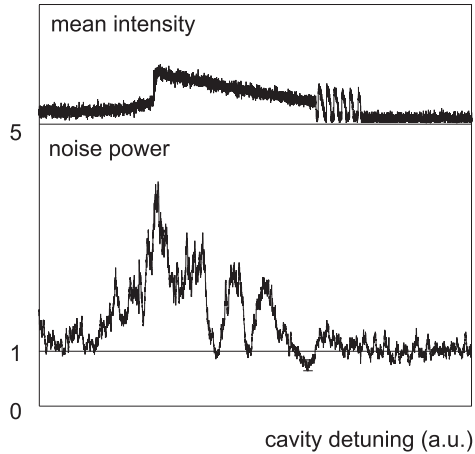,height=6cm}}
\caption{Noise signal (measured at a fixed frequency of 5MHz) 
and mean intensity
taken with weakly trapped
atoms, while the cavity length was swept with a
piezoelectric transducer and the local oscillator phase was modulated
rapidly. The solid line indicates the shot noise level. The error
bar is determined as in fig.3.}
\end{figure}
Indeed the two processes that can lead to bistability, saturation of 
the optical transition and optical pumping, are easily
distinguished by their sign. Saturation of the atomic transition causes
a decrease of the refractive index, whereas optical
pumping increases it. Thus the
bistability curves due to the two processes have steep edges on opposite
sides. In the case shown in fig.4, the left hand side of the bistability
curve can be unambiguously attributed to optical pumping. With increasing 
intensity
of the probe beam, the dominant phenomenon becomes the
saturation of the atomic transition \cite{ref8} and the steep edge
changes sides. In this second experiment
squeezing is reproducibly observed only in the range of
parameters where the bistability curve originates from optical pumping.
However, squeezing is still linked to the saturation of the atomic
transition and the observed features can be interpreted as a consequence of
the dynamical processes that take place in the cavity. As the cavity length
is scanned, light from the probe beam enters the cavity and pumps the atoms
towards the sublevels with the highest magnetic quantum number. At that stage, saturation 
of
the atomic transition starts to take place, and this non-linearity is at the
origin of squeezing in the output field. As was shown in \cite{ref8}, the
instabilities on the right hand side of the curve (after
the region where the squeezing occurs) come from a competition between
optical pumping, which is not fully completed here, and saturation of
the atomic transition.

The experimental spectra for free atoms have been compared with the theoretical predictions
given by the two-level atom model derived from ref.\cite{ref10}. The
measured experimental parameters are included in the model, which takes into
account the Gaussian character of the beam.
The minimal and maximal quantum noise were calculated in conditions
reproducing those of the experiments, i.e. by including the variation in
time of the cooperativity as represented in fig.2. and the homodyne efficiency of $90\%$. 
The resulting
spectra are shown by the broken lines in fig.3. It can be seen that the
agreement between theory and experiment is satisfactory.

As far as the squeezing in presence of the trap is concerned, one
can calculate the value of squeezing from the same two-level atom
theory. The expected quantum noise reduction is then 20\%, in good agreement
with the observed value. Squeezing is here limited by
the lower value of the cooperativity which is on the order of $20$.

We have evaluated the potential effect of spurious noise sources in the
experiments. In the experiment with free atoms, the excess noise due to the
atomic number fluctuations in the cold atom cloud in the process of
expanding and of falling \cite{Lambrecht} was shown to have a spectrum peaked at
frequencies that are too low to affect the noise measurements at 5MHz.
Additional causes of excess noise are the repumping laser and the weak
trapping beams in the second experiment. Their effects have been evaluated
with a calculation based on the atomic number fluctuations they produce. 
In the frequency range studied, we find their excess noise to be negligible 
for the rather low powers that are used for repumping and
trapping beams.

This study shows that cold atoms provide a powerful medium for
quantum optics. The measured squeezing is higher than any value observed in
previous atomic experiments. Even higher values are expected by increasing the ratio of the 
cavity linewidth to the atomic linewidth, i.e. by going more towards the bad-cavity limit. 
One of the difficulties of our first set of
experiments, in which  the trap had to be turned off
during the noise measurements,  is avoided in the second kind of
experiments. It should be possible to go further by working with cold
atoms that do not interact at all with the trapping beams, 
for example, by using a ``dark SPOT'' set-up\cite{ketterle}.

This work was supported in part by the EC contract ESPRIT BRA 6934 and the EC HMC 
contract CHRX-CT 930114.

\end{document}